\documentclass[apj]{emulateapj}
\usepackage{apjfonts}
\usepackage{natbib}

\begin{document}

\def\mpchi{\,h^{-1}{\rm {Mpc}}}
\def\mpchii{\,h{\rm {Mpc}^{-1}}}
\def\kpchi{\,h^{-1}{\rm {kpc}}}
\def\kms{\,{\rm {km\, s^{-1}}}}
\def\msun{\,h^{-1}{\rm M_\sun}}
\def\k{\mathbf{k}}
\def\x{\mathbf{x}}
\def\q{\mathbf{q}}
\def\r{\mathbf{r}}
\def\s{\mathbf{s}}
\def\be{\begin{equation}}
\def\ee{\end{equation}}
\def\ba{\begin{eqnarray}}
\def\ea{\end{eqnarray}}

\newcommand{\rnew}{R_{\rm new}}
\newcommand{\tdyn}{$t_{\rm dyn}$}
\newcommand{\tdyne}{t_{\rm dyn}}
\newcommand{\tdf}{$t_{\rm df}$}
\newcommand{\tdfe}{t_{\rm df}}

\newcommand{\forb}{$f_{\rm orb}$}
\newcommand{\forbe}{f_{\rm orb}}
\def\sarc{$^{\prime\prime}\!\!.$}
\def\arcsec{$^{\prime\prime}$}
\def\arcmin{$^{\prime}$}
\def\degr{$^{\circ}$}
\def\seco{$^{\rm s}\!\!.$}
\def\ls{\lower 2pt \hbox{$\;\scriptscriptstyle \buildrel<\over\sim\;$}}
\def\gs{\lower 2pt \hbox{$\;\scriptscriptstyle \buildrel>\over\sim\;$}}
\def\mbh{$M_{\rm BH}$}
\def\mhalo{$M_{\rm halo}$}
\def\mhaloe{M_{\rm halo}}
\def\mctwo{$M_{\rm 200c}$}
\def\mctwoe{M_{\rm 200c}}
\def\mtwo{$M_{\rm 200}$}
\def\mtwoe{M_{\rm 200}}
\def\mstar{$M_{\rm star}$}
\def\mstare{M_{\rm star}}
\def\msune{M_{\odot}}
\def\msun{$M_{\odot}$}
\def\sis{$\sigma$}
\def\kms{$km\, s^{-1}$}
\def\vvir{$V_{\rm vir}$}
\def\vc{$V_c$}
\def\vce{V_c}
\def\vmax{$V_{\rm max}$}
\def\vmax{V_{\rm max}}
\def\msunhe{M_{\odot}/h}
\def\msunh{$M_{\odot}/h$}
\def\re{$R_e$}
\def\ree{R_e}

\title{Avoiding progenitor bias: The structural and mass evolution of Brightest Group and Cluster Galaxies in Hierarchical models since $z \lesssim 1$}

\author{Francesco Shankar\altaffilmark{1}, Stewart Buchan\altaffilmark{1}, Alessandro Rettura\altaffilmark{2,3}, Vincent R. Bouillot\altaffilmark{4}, Jorge Moreno\altaffilmark{5,6,7}, Rossella Licitra\altaffilmark{8}, Mariangela Bernardi\altaffilmark{9}, Marc Huertas-Company\altaffilmark{8,10}, Simona Mei\altaffilmark{8,10,11},  Bego\~{n}a Ascaso\altaffilmark{8}, Ravi Sheth\altaffilmark{9,12}, Lauriane Delaye\altaffilmark{8}, Anand Raichoor\altaffilmark{8}}
\altaffiltext{1}{School of Physics and Astronomy, University of Southampton, Southampton SO17 1BJ, UK; F.Shankar@soton.ac.uk}
\altaffiltext{2}{Jet Propulsion Laboratory, California Institute of Technology,
MS 169-234, Pasadena, CA 91109, USA}
\altaffiltext{3}{Department of Astronomy, California Institute of Technology, MS 249-17, Pasadena, CA 91125, USA}
\altaffiltext{4}{Centre for Astrophysics, Cosmology \& Gravitation, Department of Mathematics \& Applied Mathematics,
University of Cape Town, Cape Town 7701, South Africa}
\altaffiltext{5}{Department of Physics and Astronomy, University of Victoria, Victoria, British Columbia, V8P 1A1, Canada}
\altaffiltext{6}{CITA National Fellow}
\altaffiltext{7}{Department of Physics and Astronomy, California State Polytechnic University, Pomona, CA 91768, USA}
\altaffiltext{8}{GEPI, Observatoire de Paris, CNRS, Universit\'e Paris
Diderot, Paris Sciences et Lettres (PSL), 61, Avenue de l'Observatoire 75014, Paris  France}
\altaffiltext{9}{Department of Physics and Astronomy, University of Pennsylvania, 209
South 33rd St, Philadelphia, PA 19104}
\altaffiltext{10}{Universit\'{e} Paris Denis Diderot, Universit\'e Paris Sorbonne Cit\'e, 75205 Paris Cedex
  13, France}
\altaffiltext{11}{California Institute of Technology, Pasadena, CA 91125, USA}
\altaffiltext{12}{International Center for Theoretical Physics, 34151 Trieste, Italy}

\begin{abstract}
The mass and structural evolution of massive galaxies is one of the hottest topics in galaxy formation. This
is because it may reveal invaluable insights into the still debated evolutionary processes governing the growth
and assembly of spheroids. However, direct comparison between models and observations is usually prevented
by the so-called "progenitor bias", i.e., new galaxies entering the observational selection at later epochs, thus
eluding a precise study of how pre-existing galaxies actually evolve in size. To limit this effect, we here gather data on high-redshift brightest group and cluster galaxies, evolve their (mean) host halo masses down to $z=0$ along their main progenitors, and assign as their ``descendants'' local SDSS central galaxies matched in host halo mass. At face value, the comparison between high redshift and local data suggests a noticeable increase in stellar mass of a factor of $\gtrsim 2$ since $z \sim 1$, and of $\gtrsim 2.5$ in mean effective radius. We then compare the inferred stellar mass and size growth with those predicted by hierarchical models for central galaxies, selected at high redshifts to closely match the halo and stellar mass bins as in the data. Only hierarchical models characterized by very limited satellite stellar stripping and parabolic orbits are capable of broadly reproducing the stellar mass and size increase of a factor $\sim 2-4$ observed in cluster galaxies since $z \sim 1$. The predicted, average (major) merger rate since $z\sim 1$ is in good agreement with the latest observational estimates.
\end{abstract}

\keywords{cosmology: theory -- galaxies: statistics -- galaxies: evolution}

%\begin{doublespace}

\section{Introduction}
\label{sec|intro}

The size evolution of massive, bulge-dominated galaxies has now turned into one of the hottest topics
in galaxy formation \citep[e.g.,][]{Trujillo07}. While there are promising ideas on how this evolution actually occurred, its exact nature is still debated.

One of the main limitations is that samples usually study galaxy evolution at fixed stellar mass,
a procedure that inevitably includes contributions from pre-existing galaxies in the sample evolving in time,
as well as new galaxies entering the selection at later times. As emphasized by several authors \citep[][]{Hop08FP,Carollo13}, the impact by newcomers on the global structural evolution of massive galaxies, usually termed as ``progenitor bias'' \citep[e.g.,][]{van96,Saglia10}, might in fact even be \emph{the} dominant factor controlling the inferred structural evolution.
There is indeed evidence for a significant increase in the number density of massive early-type galaxies with cosmic time at least since $z\lesssim 2$ \citep[e.g.,][]{Buitrago13,Huertas13a}.

To overcome progenitor bias, there have been mainly three methods proposed in the literature. By assuming number density conservation in the stellar mass function, \citet{VanDokkum10} attempted, on statistical grounds, to trace back what is the expected average evolution along the progenitors. This methodology, now widely applied from intermediate-mass \citep{Patel13,Papovich14}, to very massive galaxies \citep[e.g.,][]{Marchesini14,Owns14}, clearly presents limitations, such as the mass-dependent specific star formation rate distribution
and the impact of mergers, which could potentially alter the galaxy rank-ordering in time \citep[e.g.,][]{VanDokkum10,Behroozi13c}.
Another possibility is based on applying some ``age filtering'' \citep[e.g.,][]{Cimatti12}, i.e.,
by considering only galaxies having ages compatible with the cosmic time passed from high to low redshifts \citep[see also][]{Keating14}.
Both of the former techniques are acceptable ones to grasp some initial insights into the admittedly challenging issue of empirically defining patterns along the elusive progenitor-descendant evolutionary tracks.

A third approach \citep[e.g.,][]{Lidman12,Lin13}, which is the one followed, refined, and extended in this work, has been to focus only on a specific class of galaxies, namely the Brightest Cluster Galaxies (BCGs). The idea is to start at high redshift with BCGs matched in stellar and halo mass to a given observational sample, calculate their host cluster mass growth down to $z=0$ as expected from hierarchical models of structure formation, and finally infer their average stellar mass growth by selecting as descendants those
galaxies matched in present-day cluster mass via some empirical stellar mass-halo mass relation.

The aim of this work is to extend this methodology to specifically include \emph{sizes}. To this purpose, we perform a close comparison between state-of-the-art hierarchical models and a number of measurements of the mass and size of central galaxies in low and high-redshift groups and clusters.

This is one of the few attempts to probe mass \emph{and} structural evolution avoiding progenitor bias in both the models and the data. Moreover, BCGs are the ideal systems for probing structural evolution \citep[e.g.,][]{Edwards12}.
Since the most massive dark matter halos assembled their mass recently by major mergers, it is likely that the structural parameters of the most massive galaxies at their centers are also affected.

\section{Data}
\label{sec|Data}

This work makes use of a cluster sample drawn by \citet{Lidman12} in the redshift interval $0.85<z<1.63$ from the SpARCS survey \citep{Muzzin09,Wilson09,Muzzin12}. All 12 clusters were discovered searching for overdensities of red galaxies using a combination of {\it Spitzer} IRAC and ground-based z-band images. The BCG sizes adopted here are derived by Rettura et al. (2015) using GIM2D, a fitting algorithm for parameterized two-dimensional modeling of surface brightness distribution \citep{Simard98}, to fit each galaxy light distribution by adopting a \citet{Sersic63} profile. Halo masses in SpARCS,
inferred from line-of-sight velocity dispersions \citep[e.g.,][]{Demarco10,Ascaso13}, span two decades in cluster mass $0.043\times 10^{15} < \mhaloe/\msune < 2.629\times 10^{15}$ (Wilson et al., in prep.).

We complement this sample with data on groups at $z \sim 1$ from COSMOS \citep[][and references therein]{Finoguenov07,George11,Huertas13a}. These have halo mass measurements derived from weak lensing,
and galaxies by single  S\'{e}rsic profiles. The galaxies used here have been strictly selected to have a probability of being massive ellipticals with PROBEll$>0.5$, based on the Bayesian automated morphological classifier by \citet{Huertas11}, though we stress
that none of our results relies on such a specific selection. As extensively discussed by \citet{Huertas13a}, the resulting subsample of galaxies have high S\'{e}rsic indices with $n\gtrsim 5$.

At $z<0.2$ we use the \citet{Bernardi13,Bernardi12} galaxy sample extracted from
the \citet{Meert14} database which has new and updated photometric measurements
of the Sloan Digital Sky Survey (SDSS)-DR7 main galaxy sample (specifically improved sky subtraction and different profile fitting).
The host halos masses have been assigned by matching the \citet{Bernardi13} galaxy sample to the \citet{Yang07} halo catalog. In particular, we recomputed the total stellar mass of each group ($M^*_{gr}$) in the Yang et al. catalog (using the Meert et al. updated photometry), and rank-ordered the halo mass to this updated $M^*_{gr}$.

For completeness, we also assign galaxies to halo masses using the recent direct abundance matching results from \citet{Shankar14b}, based on the revised \citet{Bernardi13} stellar mass function and a negligible scatter in stellar mass at fixed halo mass \citep[see also][and references therein]{Rodriguez14}.

\section{Methodology}
\label{sec|Methodology}

Our methodology can be summarized in the following steps:
\begin{itemize}
\item Within our numerical dark matter catalogs, we select
those halos that fall in the same bin of halo mass of our $z \sim 1$ data (Sect.~\ref{sec|Data}).
\item Among these halos we select those that host a central galaxy within the range
of stellar masses measured in the data.
\item We follow the merger trees of the latter subsample of halos along their main progenitor branches down to $z\sim 0$.
\item In the local Universe we then select within the SDSS
catalog (cfr. Sect.~\ref{sec|Data}), those central galaxies that share the same mean
host halo mass as the descendants of our $z \sim 1$ halos.
\item The comparison between ``descendent'' and ``progenitor'' systems linked by their main progenitors branches,
informs us on the possible mean growth histories undergone by our $z \sim 1$ central galaxies.
\item We finally compare the predictions of state-of-the-art semi-empirical and semi-analytic hierarchical
galaxy evolution models for the growth in \emph{mean} stellar mass and
effective radius with the ones inferred by our combined $z\sim 1$ and $z<0.3$ data.
\end{itemize}

\section{Models}
\label{sec|Models}

\begin{figure*}
    \center{\includegraphics[width=15truecm]{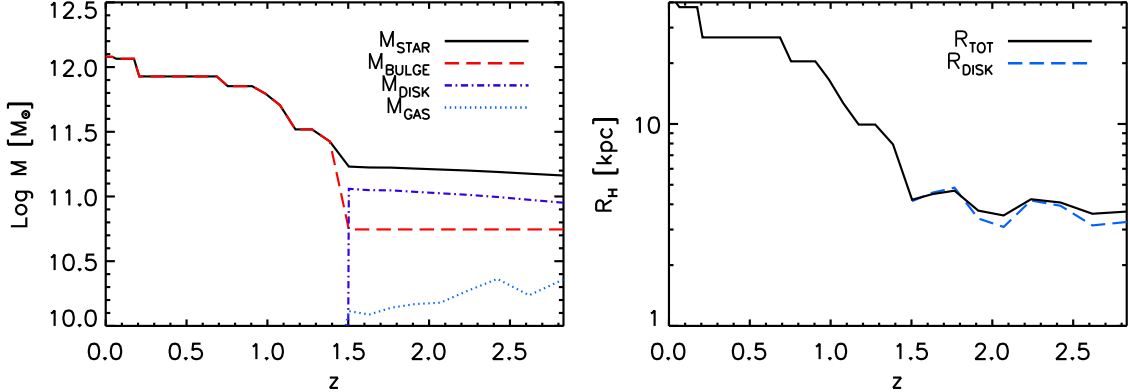}}
    \caption{\emph{Left}: Predicted mass evolution in the total, bulge, disk, and gas components, as labeled, for a central galaxy in a host dark matter halo of mass $\mhaloe \sim 1.5\times 10^{15} \msune$ at $z=0$. \emph{Right}: Corresponding evolution in the half-mass radius of the bulge and disk components; the \emph{solid}, \emph{black} line marks the mass-average of the full . The galaxy undergoes a major merger at $z \sim 0.9$ that converts it from a gas-rich disk to a gas-poor elliptical.}
    \label{fig|MergerHistoriesSingleGalx}
\end{figure*}

In the following, we take as a reference the semi-empirical model developed and extensively discussed by \citet{Shankar14}, based on the general lines exposed by \citet{Hop08FP} and \citet{Zavala12}. We here recall its main ingredients.

The model is built on top of numerical merger trees extracted from the Millennium I Simulation\footnote{By adapting the \citet{Shankar14} basic routines to also run on analytic merger trees, mainly characterized by lower values of $\sigma_8=0.8$, instead of $\sigma_8=0.9$ of the Millennium Simulation, we have checked that our main results do not depend on the exact underlying cosmology.}
\citep{Springel05}. Central galaxies, associated with the main progenitor branch of the tree, are initially defined as disks at early epochs. All their mass and structural properties are assigned via empirical, time-dependent
relations, more significantly the stellar mass-halo mass relation, taken from \citet{Moster13},
the disk radius-stellar mass relation, which we write as \citep[e.g.,][]{Shen03,Hop08FP,Somerville08}
\begin{equation}
R_{\rm disk}=\frac{R_0}{(1+z)^{-0.4}}\mstare^{k}\left(1+\frac{\mstare}{3.98\times 10^{10}\, \msune}\right)^{p-k}
    \label{eq|rdisk}
\end{equation}
with $R_0$=0.1, $k=0.14$, $p=0.39$,
the specific star formation rate-stellar mass relation
\citep[][]{Karim11}, and the mass- and time-dependent gas fraction as given by \citet{Stewart09}.

Central galaxies are re-initialized at each timestep during the evolution and can gradually transform their morphology via mergers with satellites. Satellite galaxies are those associated with each dark matter branch merging with the main progenitor.
When galaxies become satellites in larger haloes, they are assigned a dynamical friction time-scale
for final coalescence with the central galaxy taken from \citet{McCavana12}.
They are also assigned all the mass and structural properties of a central galaxy in a typical halo, within the Monte Carlo catalog,
of the same mass at the time of infall. The evolution of satellites
after infall takes into account extra mass and size growth in the disks from the residual
gas, and may also include gas and stellar stripping in some models.

When a merger between two progenitor galaxies of masses $M_1$ and $M_2$ is major ($M_2/M_1>0.3$), it is assumed that a single elliptical galaxy is produced, and any gas present in the disks of the merging
galaxies is converted into stars in a burst.
In a minor merger ($M_2/M_1<0.3$), all the stars of the accreted satellite are added to the bulge of the central
galaxy, while the gas is added to the main gas disk. New stars are also formed in a merger
following the collisional starburst model by \citet{Somerville01}, where a fraction
\begin{equation}
e_{\rm burst} = 0.56 \left(\frac{M_2}{M_1}\right)^{0.7}
\label{eq|eburst}
\end{equation}
of the cold gas of the progenitors is converted into stars.

The size of the remnant is computed from the energy conservation between the sum of the self-binding
energies of the progenitor galaxies, and that of the remnant \citep{Cole00},
\begin{equation}
\frac{(M_1+M_2)^2}{\rnew}=\frac{M_1^2}{R_1}+\frac{M_2^2}{R_2}+\frac{f_{\rm orb}}{c}\frac{M_1M_2}{R_1+R_2}\, ,
\label{eq|sizegeneral}
\end{equation}
where $M_i$, $R_i$, are, respectively, the total masses and half-mass radii of the merging
galaxies, and the form factor $c \sim 0.5$. The factor $f_{\rm orb}$ parameterizes the average orbital energy of the merging systems, ranging from zero to unity for parabolic and circular orbits, respectively.
Eq.~\ref{eq|sizegeneral} can further be adapted to include any type of
energy loss such as gas dissipation \citep[e.g.,][]{Hop08FP,Covington11,Porter14,Shankar14}.

Our semi-empirical model does not include
other ``in-situ'' processes, such as major disk instabilities \citep[e.g.,][]{Bower06} or disk regrowth after a major merger \citep[e.g.,][]{Hammer09}. We note, however, that extensive semi-analytic models have proven
disk regrowth to play a significant role mainly in intermediate-mass galaxies \citep{DeLucia11,Wilman13}, below
the stellar mass threshold considered in this work. We also note that our
massive ellipticals, remnants of a major merger, are usually devoid of any left-over gas,
which is mostly consumed in the star formation
burst (Eq.~\ref{eq|eburst}; see also Fig.~\ref{fig|MergerHistoriesSingleGalx}).
In some sporadic cases, some later gas-rich minor mergers may
be able to re-grow a light stellar disk, but this has negligible impact
on any of our results.

Fig.~\ref{fig|MergerHistoriesSingleGalx} shows an example of the mass and structural evolution of a central galaxy in one of the most massive halos in our catalog, with mass $\mhaloe\sim 1.2\times 10^{15} \msune$ at $z=0$. The galaxy starts as a disk at $z \sim 3$ (left panel), with a small
central bulge built via minor mergers, reduces its gas reservoir, and finally fully transforms its morphology, turning, after a major merger at $z\sim 1.5$, into an elliptical with no residual gas. The subsequent evolution is mainly driven by dry minor and major mergers, with its half-mass radius increasing by a factor of $\sim 10$ since its first major-merger event (right panel).

For completeness, in the following we will also compare our data with the publicly available\footnote{http://www.g-vo.org/MyMillennium3} hierarchical models by \citet{Guo11} and \citet{Henri14}. Specifically, we have analyzed several thousands of merger trees extracted from their online catalogs, i.e., those contributing to all central galaxies with final stellar mass $\mstare \gtrsim 3\times 10^{11} \, \msune$.
We discuss the relevant specifications and predictions of such models in Sect.~\ref{subsec|ModelVsData}.

\section{Results}
\label{sec|results}

\subsection{The overall size evolution of BCGs}
\label{subsec|MedianMassSize}

\begin{figure*}
    \center{\includegraphics[width=15truecm]{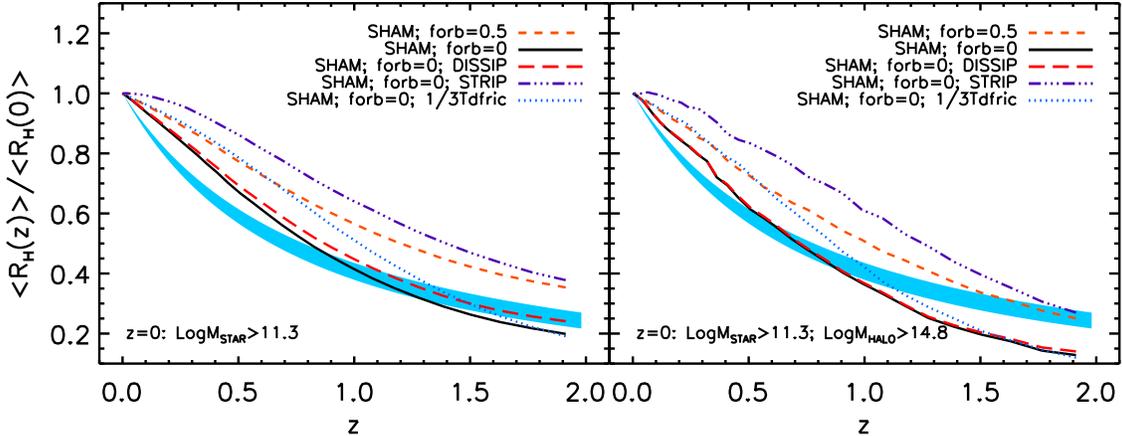}}
     \caption{Predicted mean size evolution, normalized to the value at $z=0$ for different models, as labelled, for galaxies that today have a stellar mass $\log \mstare > 11.3\, \msune$. The \emph{left} panel considers averages all galaxies in all environments, while the \emph{right} panel only considers galaxies in cluster environments, with $\log \mhaloe > 14.5\, \msune$, which have a stronger size evolution. Models with significant stripping in the infalling satellites, larger orbital energies, and lower dynamical frictions timescales are disfavoured by the derived observational estimates by, e.g., \citet{VanDokkum10} (\emph{cyan} region).}
    \label{fig|SizeEvolFixedMstar}
\end{figure*}

We start in Fig.~\ref{fig|SizeEvolFixedMstar} by reporting the observational results by \citet[][cyan region]{VanDokkum10}. They fitted the mean size evolution of $z=0$ galaxies with $\mstare \gtrsim 2\times 10^{11}\, \msune$ along the ``progenitors'', identified via number density conservation (Sect.~\ref{sec|intro}), as $R_e(z)\propto (1+z)^{1.3}$ \citep[see also, e.g.,][]{Cimatti12}.
Their results are compared to the main semi-empirical model realizations, as labeled, developed by \citet{Shankar14} for galaxies selected to have the same cut in stellar mass at $z=0$, and followed in their mean half-mass radius growth along their main progenitor branches.

All models point to an increase in median size
of a factor of $\lesssim 2.5$ since $z\sim 1$, in broad agreement with the data.
However, not all models perform equally well.
Models characterized by high orbital energies ($\forbe \gtrsim 0.5$; dashed, orange lines) have already been shown
to predict shallower size-mass relations compared to data \citep{Shankar13,Shankar14}.
Fig.~\ref{fig|SizeEvolFixedMstar} extends these results by showing that the
latter type of models also induce a weaker size evolution with respect to the data.
As already pointed out by \citet{Shankar13}, high-resolution cosmological simulations
aimed at studying the orbital parameters of major mergers of cold dark matter halos,
find indeed that a large fraction of encounters are nearly parabolic \citep{KhochfarBurkert06}.

Stellar stripping in the infalling satellites is included in one variation of the
reference model by following the empirically motivated mass-dependent stellar and gas stripping,
parameterized as \citep{Cattaneo11}
\begin{equation}
{\rm F_{\rm strip}}=(1-\eta)^{\tau} \, ,
\label{eq|stripping}
\end{equation}
with $\tau=\tdfe/\tdyne$ the ratio between the dynamical friction and dynamical timescales.
It was discussed by \citet{Shankar14} that a value of $\tau=0.25$ allows to
faithfully match the local \citet{Moster13} stellar mass-halo mass relation.
Without stripping, the resulting $z=0$ stellar mass-halo mass relation would end-up being
steeper, in better agreement with other, more recent determinations of the stellar mass-halo mass relation based on the \citet{Bernardi13} stellar mass function and BCG measurements (see, e.g., \citealt{Kravtsov14} and \citealt{Shankar14b}).

In \citet{Shankar14} we showed that models with stellar stripping tend
to fall short in matching the local size distributions especially at the high-mass end,
mainly due to the merging satellites turning too compact to promote efficient size growth.
Here we extend such result to higher redshifts, emphasizing that the same models with stellar stripping
(triple dot-dashed, purple lines in Fig.~\ref{fig|SizeEvolFixedMstar}) tend, as expected, to generate a size evolution too weak with respect to what empirically inferred.

Models with significantly lower dynamical friction timescales, specifically $1/3$ of the ones used in the reference model (dotted, cyan lines), increase the number of mergers on the central galaxy and have been shown to be disfavored with respect to the local data in terms of their large intrinsic scatter and more pronounced environmental dependence \citep{Shankar14}. These models also tend to produce a faster growth with respect to observational results, especially at $z>1$, though per se deep data are still too sparse to be able to firmly rule out this model, as further discussed below. Gas dissipation in major mergers (long dashed, red lines) does not significantly alter the size evolution along the progenitors \citep[see also][]{Shankar13}.

In the following, we will adopt the $\forbe=0$, no-stripping, and no-dissipation semi-empirical model as our reference one, as it performs better against local data \citep{Shankar14}, and predicts reasonable degrees of size evolution.

Probing size evolution at fixed stellar mass, besides the effects of progenitor bias, could also mix different environments at different epochs \citep[][]{Valentinuzzi10a,Poggianti13}, preventing a clear understanding of the true role of environment in size evolution. In fact, as evident from the right panel of Fig.~\ref{fig|SizeEvolFixedMstar}, the subsample of central galaxies
specifically hosted by massive clusters above $\mhaloe \gtrsim 6\times 10^{14}\, \msune$ at $z=0$, share quite similar mean evolutionary histories with respect to all galaxies of similar stellar mass, at least at $z\lesssim 1$, while they tend to have an accelerated evolution at $z>1$. Although environment may not play a dominant role in determining the degree of size evolution \citep{Huertas13b,Shankar14}, it may still accelerate the growth of galaxies living within the depth of large potential wells \citep[][]{Fassbender14,Mei14}.
Signs of an accelerated evolution in (proto)clusters is now becoming more and more evident in high redshift data from different groups \citep[e.g.,][and references therein]{Lani13,Delaye14}, and we will reserve separate work on this and other related issues.

\subsection{The mean and median growth of host halos}
\label{subsec|MeanMedianGrowths}

The comparison between models and data attempted in the previous section is only a very approximate one. The technique of number conservation is in fact only a rough approximation for tracking progenitors. Indeed, we verified with the aid of the full \citet{Guo11} merger tree catalog of massive galaxies, that selecting at fixed number density tends to induce a somewhat weaker size evolution than the true one. We thus proceed below with a more accurate model-to-data comparison based on exploiting the high and low-$z$ BCG data listed in Sect.~\ref{sec|Data}, following the methodology described in Sect.~\ref{sec|Methodology}.

\begin{figure*}
    \center{\includegraphics[width=15truecm]{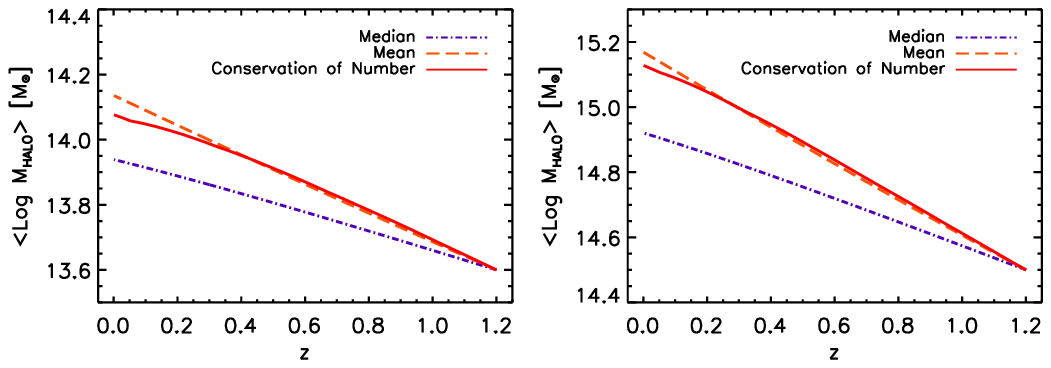}}
    \caption{Analytic fits by \citet{Fak10} to the mean (\emph{orange}, \emph{long-dashed}) and median (\emph{purple}, \emph{long-dashed}) accretion rates for halos with mass of $\log \mhaloe/\msune=13.7$ (\emph{left}) and $\log \mhaloe/\msune=14.5$ (\emph{right}) at $z=1.2$. The \emph{solid}, \emph{red} lines show the evolution of the (minimum) halo mass for which the number density equals the one at $z=1.2$ of halos with mass $\log \mhaloe/\msune=13.7,14.5$. Mean accretion rates more closely track the mass evolution corresponding to the conservation of cumulative number density.}
    \label{fig|MeanMedianGrowthRates}   
\end{figure*}

However, before proceeding with such a comparison between theory and observations, it is essential to point out the implied differences between median and mean dark matter halo growths. Fig.~\ref{fig|MeanMedianGrowthRates} reports
the mean (orange, long-dashed) and median (purple, long-dashed) halo mass growth of halos with mass $\log \mhaloe/\msune=13.7$ (left) and $\log \mhaloe/\msune=14.5$ (right) from $z=1.2$ down to $z=0$. These curves are obtained by integrating the respective mean and median accretion rates
presented by \citet{Fak10}, derived from direct fits to the Millennium simulations. Mean accretion rates are at all redshifts a factor of about two higher than median ones, thus implying a much more prominent mass growth.

The solid, red lines in Fig.~\ref{fig|MeanMedianGrowthRates} mark instead the evolution of the (minimum) halo mass for which the halo cumulative number density at all subsequent times equals the cumulative number density above a halo mass of $\log \mhaloe/\msune \ge 13.7,14.5$ at $z=1.2$ (left and right panel, respectively). The latter evolutionary tracks nicely follow the \emph{mean} accretion rates down to $z\sim 0$, thus proving that mean accretion rates are better suited to track the average mass evolution of a subsample of halos from high to low redshifts. In the following we will thus always adopt \emph{mean} quantities to describe the evolution of galactic properties.

\subsection{Specific comparisons between model predictions and observational data}
\label{subsec|ModelVsData}

\begin{figure*}
    \center{\includegraphics[width=18truecm]{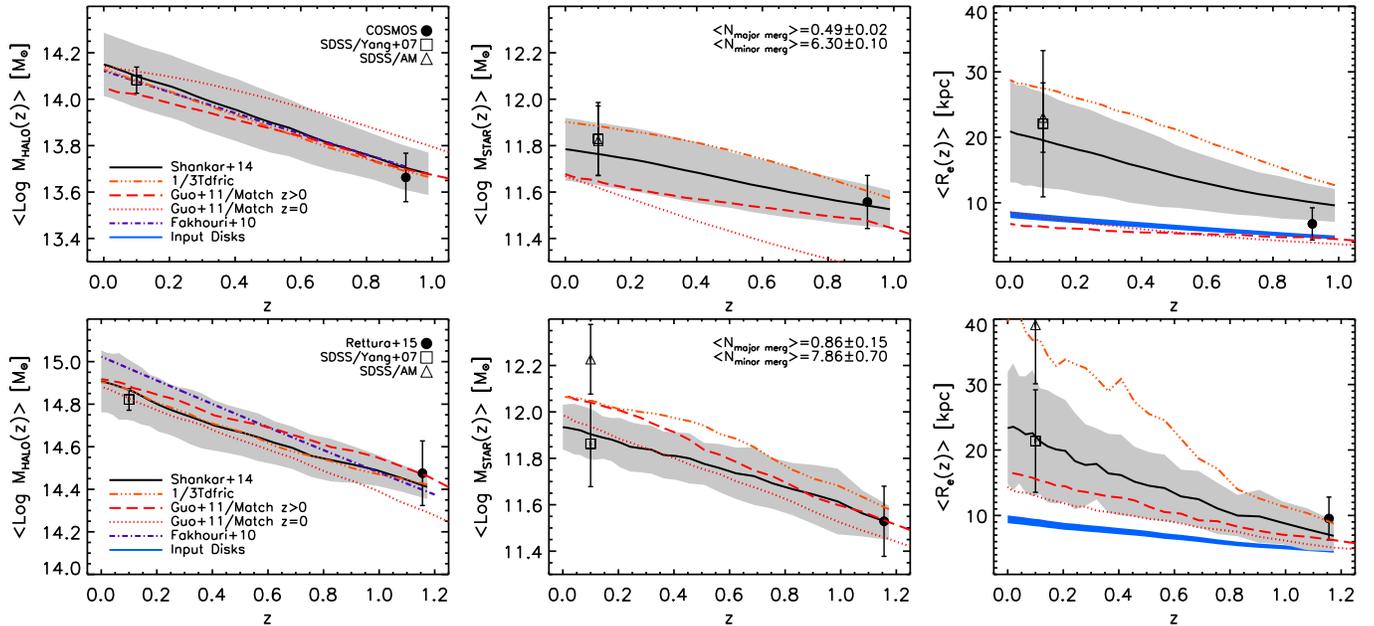}}
    \caption{Mean halo mass (\emph{left}), stellar mass (\emph{middle}), and projected half-mass/light radius (\emph{right}) evolutions predicted by the reference semi-empirical model (\emph{solid} lines with \emph{gray bands}), the semi-empirical model with lower dynamical friction timescales (\emph{triple dot-dashed}, \emph{orange} lines), and the semi-analytic model of \citet{Guo11} (\emph{red long-dashed} and \emph{dotted} lines). The \emph{upper} panels compare with the data by COSMOS \citep{Huertas13a} (\emph{filled circles}) and SDSS local data, while the \emph{lower} ones with the recent data by SpARCS (Rettura et al. 2015, in prep.) and SDSS. Local SDSS data on stellar masses and sizes (\emph{middle} and \emph{right} panels, respectively) extracted from the \citep{Yang07} halo catalog matched to the \citet{Bernardi12} SDSS DR7 sample are shown with \emph{open squares}, while those derived from direct abundance matching \citep{Shankar14b} are shown with \emph{open triangles}. Models broadly reproduce the data, especially in group-sized halos. Lower dynamical friction timescale models tend to provide faster growth tracks, in apparent better agreement with massive cluster environments.}
    \label{fig|SizeEvolutionCosmosRettura}
\end{figure*}

Fig.~\ref{fig|SizeEvolutionCosmosRettura} compares the halo mass, stellar mass, and size evolutionary tracks predicted by hierarchical models to observational data. The filled circles refer to high-redshift data from COSMOS at $z \sim 0.9$ (upper panels) and SpARCS at $z \sim 1.2$ (lower panels). In passing, we note that the $z \sim 1$ stellar and host halo masses in our reference data are,
within the nominal scatter of $\sim 0.15$ dex \citep[e.g.,][]{Rodriguez14,Shankar14b}, in line with what predicted by abundance matching models at these redshifts \citep[e.g.,][]{Moster10,Shankar14b}.

Local data from SDSS are marked by open symbols. The stellar masses (middle panels) and corresponding effective radii (right panels) inferred from the \citet{Yang07} halo catalog matched to the \citet{Bernardi12} SDSS DR7 sample, are shown with open squares, while those derived from direct abundance matching \citep{Shankar14b} are shown with open triangles. The latter, in line with other groups \citep{Gonzalez13,Kravtsov14}, predict at fixed halo mass stellar masses (and thus related sizes) in very good agreement with those derived from the \citet{Yang07} halo catalog for halo masses $\mhaloe \lesssim 2\times 10^{14} \, \msune$ (upper panels). For more massive host halos, revised abundance matching models \citep{Kravtsov14,Shankar14b} predict stellar masses systematically higher by a factor of $\sim 2$, and consequently effective radii higher by a factor of $\sim 1.6$.

At face value, the simple comparison between the high-$z$ (filled circles) and local SDSS data (open symbols), irrespective of the exact stellar mass-halo mass mapping, suggests a significant growth of a factor of $\gtrsim 2$ in mean stellar mass since $z\sim 1.0$, and a comparable increase of a factor of $\gtrsim 2.5$ in mean effective radius for central galaxies in both the group and cluster environments. These trends support an average growth rate of $\Delta R_e \propto \Delta \mstare^{\alpha}$ with $\alpha \gtrsim 1.3$, thus stronger than a model based on only major mergers ($\alpha=1$), indicative of the possibly significant role played by minor mergers in the global structural growth.

We now proceed in checking if the latter, observationally-based findings reported in Fig.~\ref{fig|SizeEvolutionCosmosRettura}, are in line with the predictions of state-of-the-art, galaxy evolution hierarchical models. Fig.~\ref{fig|SizeEvolutionCosmosRettura} is one of the first examples in the literature in which a careful comparison is carried out between models and data to probe the parallel evolution of central, massive galaxies in halo mass (left panels), stellar mass (central panels), and effective radius (right panels). In this section we detail the outputs of the semi-empirical model, while we devote the next section to the outcomes of advanced semi-analytic models. As outlined in Sect.~\ref{sec|Methodology},
for a careful comparison between models and data we first select in our mock catalogs the progenitor host halos that at $z\sim 0.9$ and $z\sim 1.2$ have masses comparable to the clusters used by \citet[][top panels]{Huertas13a} and Rettura et al. (2015; bottom panels), respectively. Among these progenitors we then select those halos that have a central galaxy with stellar mass comparable to those inferred for the central group and cluster galaxies in the reference data.
The main progenitor branches competing to the selected halos are then followed forward in time, and at each timestep we compute their mean host masses, stellar masses, and effective radii. The theoretical mean and $1~\sigma$ uncertainties are shown as solid lines and gray areas, respectively.

We first note that the mean halo masses of our selected halos (solid lines in the left panels of Fig.~\ref{fig|SizeEvolutionCosmosRettura}) evolve as predicted by the \citet{Fak10} analytic fits, within an accuracy of $\lesssim 0.05-0.1$ dex (dot-dashed, purple lines, left panels). The mean stellar mass of their central galaxies instead is predicted in the reference semi-empirical model (solid lines in the left panels of Fig.~\ref{fig|SizeEvolutionCosmosRettura}) to evolve by a factor of $\sim 2$ since $z\lesssim 1$, irrespective of the exact environment, either a group or a cluster. This growth is broadly consistent with both observational data sets at $z \sim 1$ and $z \sim 0.1$, especially for lower-mass halos (upper panels), where the local constraints from abundance matching and group catalogs are in very close agreement.

The size growth is analogously similar in the two environments, possibly predicted to be marginally weaker in the group environment.
To appropriately compare with the data we follow the same approach pursued in \citet{Shankar14}. We first assign to each mock galaxy a S\'{e}rsic index $n$ from the empirical $n-\mstare$ relation calibrated at $z=0$ \citep{Huertas13a}, and then computed the corresponding 3D-to-2D size correction from the tabulated values of \citet{Prugniel97}.
We assume constant S\'{e}rsic distribution for bulge-dominated galaxies since $z\sim 1$ \citep{Huertas13a,Ascaso13}, though we verified that even allowing for significant evolution, e.g., $n \propto (1+z)^{-0.95}$ \citep{VanDokkum10}, yields quite similar results, with only a slightly faster size evolution at $z>0.5$ in the latter case, but still well consistent within the $1~\sigma$ uncertainties.
Our model results are in broad agreement with the numerical outputs by, e.g., \citet{Laporte13}, specifically targeting the evolution of BCGs, and also with those by the \citet{Oser12}, though the latter were mainly focused on lower mass halos than the ones considered here.

We note that at the massive cluster scale, for host halos above $\mhaloe \gtrsim 5\times 10^{14}\, \msune$, the reference model tends to fall short in fully reproducing the local mean stellar mass and size as predicted by the most recent abundance matching results (open triangles). A semi-empirical model characterized by lower dynamical friction timescales (triple dot-dashed, orange lines), tends to better align with the latter sets of local data.
However, a fine-tuning should be necessary for the latter model to be fully applicable, in order not to overproduce the stringent constraints of low scatter in the local scaling relation between effective radius and stellar mass \citep{Shankar14}.
If future measurements will fully confirm the abundance matching results, it may signal the need for some revisions in key assumptions behind present hierarchical models, or even the need for alternative processes beyond only mergers contributing to shaping the size growth of massive galaxies as a function of time and environment \citep[e.g.,][and references therein]{Fan10,Shankar13,Stringer14}.

The predicted sizes of centrals in groups is also systematically larger by a factor of $\sim 1.3$ than what actually measured in the data, especially for the semi-empirical model characterized by lower dynamical friction timescales.
This discrepancy can be only partially explained by an empirical underestimation of the outer light profiles, as the stellar masses are well matched to the ones in the models, and group environments should be less prone to significant biases in sky background subtractions, at least with respect to clusters \citep{Bernardi13}.
From a purely theoretical point of view, this relatively mild discrepancy could be attributed to the progenitor disk sizes being too large at $z \sim 1$. The blue contours in the right panels of Fig.~\ref{fig|SizeEvolutionCosmosRettura} mark the input disk sizes (Eq.~\ref{eq|rdisk}) for progenitors\footnote{We note that disk sizes in Figs.~\ref{fig|SizeEvolutionCosmosRettura} and \ref{fig|SizeEvolutionCosmosRetturaHenri} have not been corrected by any projection factor \citep[e.g.,][and references therein]{Gonzalez09}} of stellar mass 0.5 and 0.7 times the remnant stellar mass, i.e., the typical progenitor masses in a major mergers, the events that in the model are responsible for creating ellipticals (cfr. Sect.~\ref{sec|Models}). It is clear that the adopted disk sizes, although empirically motivated (Sect.~\ref{sec|Models}), are already noticeably large by $z \sim 1$ thus clearly creating remnants that will inevitably overshoot the sizes of at least some ellipticals observed at these epochs.

\subsection{Comparison with advanced Semi-Analytic Models}
\label{subsec|SAM}

\begin{figure*}
    \center{\includegraphics[width=18truecm]{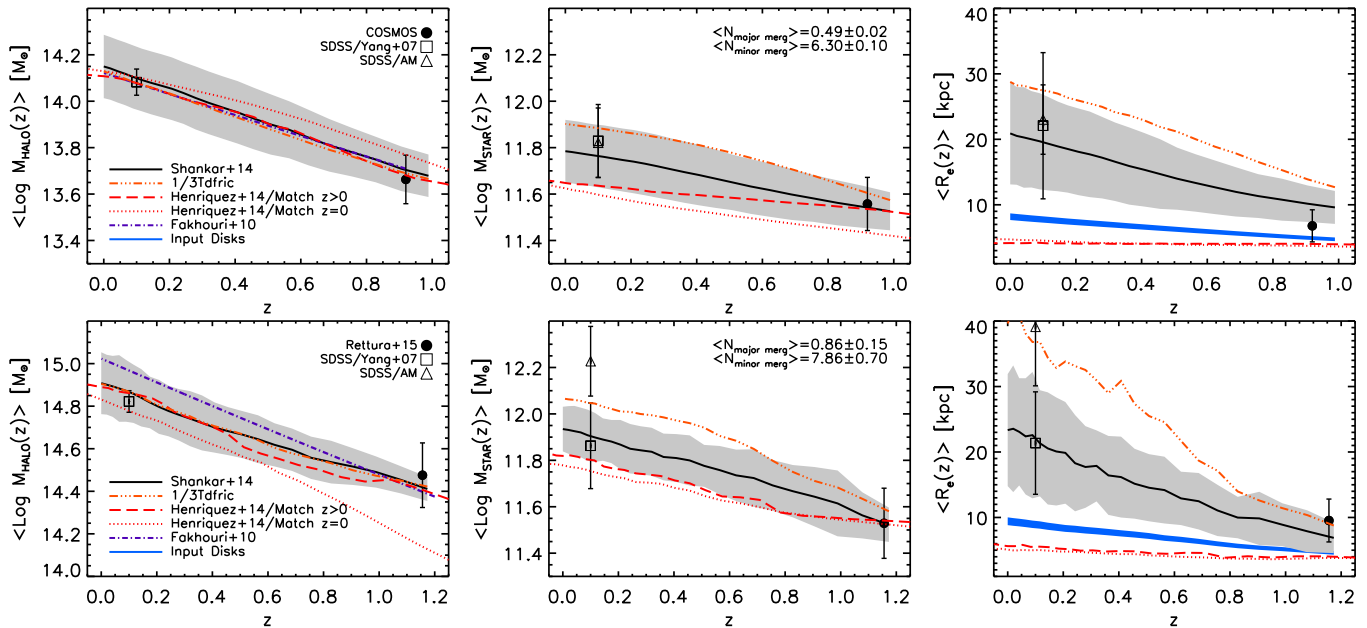}}
    \caption{Same format as Fig.~\ref{fig|SizeEvolutionCosmosRettura} but with the semi-analytic model of \citet{Guo11} replaced by the more recent version by \citet{Henri14} (\emph{red long-dashed} and \emph{dotted} lines). The predicted stellar mass and size growth in the latter model is significantly lower (see text for details).}
    \label{fig|SizeEvolutionCosmosRetturaHenri}
\end{figure*}

For completeness, in the same Fig.~\ref{fig|SizeEvolutionCosmosRettura} we also show the predictions from the \citet{Guo11} semi-analytic model (SAM). The red, dashed lines refer to the mean properties of the main progenitor galaxies selected to have halo masses similar to those in the observational samples at $z>0$, while the red, dotted lines refer to the subsample of galaxies sharing halo masses as close as possible to those of the SDSS galaxies selected as described in the previous section. It is clear that the progenitor galaxies in the SAM are hosted by dark matter halos that evolve very similarly to what predicted by the semi-empirical model. The stellar mass growth is instead predicted to be somewhat stronger than the one in the reference semi-empirical model (solid lines), thus explaining why their mean size growth is still a factor of $\sim 3$ despite using $\forbe>0$. As discussed in reference to Fig.~\ref{fig|SizeEvolFixedMstar}, models with higher values of $\forbe>0$ would be in fact expected to have less pronounced
evolution in size at fixed progenitor masses, following from Eq.~\ref{eq|sizegeneral}. The overall sizes at all redshifts in this SAM fall below the measured values.

A more recent rendition of the \citet{Guo11} model has been presented by \citet{Henri14}. The latter version differs from the previous one in several key properties. First off, they adopted the first-year Planck cosmology \citep{Planck13}, at variance with \citet{Guo11} who adopted the cosmology of the Millennium simulation, consistent to the one used in our semi-empirical model. The more relevant differences to our purposes come however from the baryonic sector. The number of mergers onto the centrals is significantly reduced in the \citet{Henri14} model
with respect to \citet{Guo11}, mainly because of two reasons. First, their flatter galaxy stellar mass function at the low-mass end, induces less numerous infalling satellites,
and second, each satellite undergoes a more efficient stellar stripping after infall.
As evident from Fig.~\ref{fig|SizeEvolutionCosmosRetturaHenri}, while the evolution of their mean dark matter host halos (left panels) remains very similar to the one predicted by \citet{Guo11}, their mean stellar mass growth is significantly reduced. Nevertheless, centrals in clusters may still undergo sufficient stellar mass growth, at least when taking the SDSS group catalog as a local reference (open square, lower-middle panel), but definitely fall short in matching the size growth (right panels). As expected, their mean effective radii in fact now evolve significantly less in the \citet{Henri14} model, at most by a factor of $\lesssim 1.5$. Clearly setting $\forbe=0$ \citep[e.g.,][]{Shankar13} and possibly reducing some of the stellar stripping in their model, should help increasing their predicted sizes and growth rates, in better agreement with the data presented in this work.
%This may be partly due to the way infalling satellites continue growing to substantially grow in stellar mass before coalescence with the central.

\section{Discussion and Conclusions}
\label{sec|discu}

We have shown that by connecting high and low redshift BCG data via evolution of their host halo masses
would imply, at face value, an increase since $z \sim 1$ of a factor $\sim 2-3$ in their mean stellar mass and $\sim 2.5-4$ in their mean effective radius.
Our results on the stellar mass evolution of BCGs are in line with previous studies by, e.g., \citet{Lidman12}, \citet{Lidman13}, and \citet{Ascaso13}.

Although we verified that our inferred mean stellar mass and size growths hold even switching to other high-redshift data sets \citep[e.g.,][]{Ascaso13,Delaye14},
still our results rely on the quality of the high redshift imaging.
Surface brightness dimming effects, varying stellar population gradients, and/or other related
spectral/photometric effects, could clearly limit the general reliability of high redshift data.
Deeper imaging, possibly accompanied by accurate dynamical measurements, will certainly be able to provide more stringent constraints on the actual stellar mass and size growths of BCGs.

At face value, we find a good match between observationally-derived stellar mass and size growth rates, and those predicted from state-of-the-art, hierarchical galaxy evolution models. In the specific, our best models predict an average number of $\sim 0.5-0.9$ major mergers and $\sim 7-8$ minor mergers since $z\sim 1$ (Fig.~\ref{fig|SizeEvolutionCosmosRettura}), with mass ratios $\mu$ in the progenitors of $0.3<\mu<1.0$ and $0.01<\mu<0.3$, respectively. In accordance to our results, in the ALHAMBRA survey \citet{Sanjuan14} recently estimated in the same redshift interval an average number of major mergers per galaxy of $N_m=0.57\pm0.05$ for red luminous galaxies (see also \citealt{Lopez12}). Our predicted average major merger rates of $R\sim (0.07-0.11)\, {\rm Gyr^{-1}}$ is also in good agreement with \citet{Lackner14} inferred from a mass completed sample of massive galaxies $R\sim 0.1\, {\rm Gyr^{-1}}$, down to $R\sim 0.04\, {\rm Gyr^{-1}}$ for quiescent galaxies alone.

\citet{Liu14} recently inferred from BCG deep, high-resolution imaging a major merger rate of $(0.55 \pm 0.27)\, {\rm Gyr^{-1}}$ at $z \sim 0.43$, concluding that present-day BCGs should have increased their stellar mass by $\sim (35 \pm 15)\%$ via major dry mergers since $z=0.6$, in good agreement with what derived from our best models. \citet{Ina14}, in analogy to our methodology, constructed from the Planck Sunyaev-Zel'dovich cluster catalog galaxy progenitor-descendant pairs of BCGs, finding an average 2-14\% growth in the redshift range 0.2-0.4 (see also \citealt{Oliva14}), in good agreement with our average growth of $\sim 10\%$ in the same redshift interval. Clustering measurements provide average merger rates of $\sim 0.024\, {\rm Gyr^{-1}}$ \citep[e.g.,][and references therein]{Wake08,White08} at $z<0.6$ somewhat lower than our estimates.

\citet{Man14} from a mass-complete sample
of massive galaxies from UltraVISTA/COSMOS, complemented with deeper, higher resolution 3DHST+CANDELS data,
have also estimated a major merger rate $R\sim 0.1\, {\rm Gyr^{-1}}$ for comparably massive galaxies,
at least since $z\sim 1.5$ (e.g., their Fig.~12). The authors suggest that
such an amount of major merging alone could be sufficient to explain the observed number density evolution for massive galaxies ($\log \mstare/\msune \ge 11.1$).
Indeed, although still debated, the combined information from current stellar mass, clustering, and weak lensing
measurements on the stellar mass and environments of intermediate-redshift BCGs tend to favor a noticeable increase in the number density of the most massive
galaxies (e.g., Buitrago et al. 2013, \citealt{Shankar14b}, and references therein).
%\citep[e.g.,][and references therein]{Buitrago13,Shankar14b}.}

However, \citet{Man14} also measure a minor merger rate of $R\sim 0.07-0.1\, {\rm Gyr^{-1}}$ that would be insufficient by itself to explain the rapid mean size evolution of a factor of $\sim 2-2.5$ since $z\sim 1$ as inferred from our data \citep[see also, e.g.,][]{Huertas13a,Man14,van14}.
However, we should note that the \citet{Man14} estimate is limited to minor mergers with a mass ratio $0.1<\mu<0.3$, while hierarchical models predict, as in the analysis presented
in this work, that extending this limit to $0.01<\mu<0.3$, increases the rate by a factor of $\gtrsim 5$ \citep[e.g.,][and references therein]{HopkinsMergers}. Such a significantly increased rate of minor mergers has a relatively lower repercussion on the increase of the stellar mass, but more effectively impacts the overall size growth \citep[e.g.,][]{Trujillo07,Naab09}.

We showed that the models with lower dynamical friction timescales, though disfavored on more general grounds because predicting too loose scaling relations at low redshifts \citep[e.g.,][]{Bower06,Shankar14}, may still be a viable solution for properly growing massive, central galaxies in cluster-sized environments. The number of predicted major mergers in the latter model is also similar to the one predicted by the reference model with longer dynamical friction timescales, though the number of minor mergers is double. Unfortunately, present data cannot still reliably constrain the number of minor mergers to distinguish among these models.

To conclude, in order to limit the effect of progenitor bias, we have gathered
a number of data on high redshift brightest group and cluster galaxies, evolved their mean host halo masses down to $z=0$ along their merger trees main progenitor branches, and assigned as their descendants local SDSS central galaxies matched in host halo masses. At face value, the direct comparison of local ``descendants'' and $z\sim 1$ ``progenitors'', suggests
an increase in mean stellar mass of a factor of $\gtrsim 2$, and a relative increase of mean effective radius of $\gtrsim 2.5$. We found that the most favored hierarchical models, with no strong stellar stripping and null orbital energies, yield a factor of $\sim 2.5-3$ growth in median size, in good agreement with what inferred from the matching between high and low redshift data, especially for the stellar mass growth of centrals in group-sized, high-redshift halos.

\vspace{1cm}

\begin{acknowledgements}
FS acknowledges partial support from a Marie Curie grant. S.M. acknowledges financial support from the Institut Universitaire de France (IUF), of which she is senior member. FS thanks Bruno Henriquez and Chervin Laporte for interesting discussions. This work is based on data obtained with the {\it Spitzer Space Telescope}, which is operated by the Jet Propulsion Lab (JPL), California Institute of Technology (Caltech), under a contract with NASA. Support was provided by NASA through contract number 1439211 issued by JPL/Caltech. VB is supported financially by the National Research Foundation of South Africa. We thank the referee for useful comments and suggestions.
\end{acknowledgements}

\bibliographystyle{yahapj}
\bibliography{../../../RefMajor_Rossella}

\label{lastpage}
\end{document}